\documentclass[12pt]{article}
\usepackage{amsmath,bm}

\renewcommand{\thefootnote}{\fnsymbol{footnote}}
\newcommand{\vev}[1]{{\langle{#1}\rangle}}

\begin{document}

\title{
\begin{flushright}
\begin{minipage}{0.2\linewidth}
\normalsize
WU-HEP-12-01 \\
KUNS-2394 \\*[50pt]
\end{minipage}
\end{flushright}
{\Large \bf 
Superfield description of 10D SYM theory 
with magnetized extra dimensions
\\*[20pt]}}

\author{Hiroyuki~Abe$^{1,}$\footnote{
E-mail address: abe@waseda.jp}, \ 
Tatsuo~Kobayashi$^{2,}$\footnote{
E-mail address: kobayash@gauge.scphys.kyoto-u.ac.jp}, \ 
Hiroshi~Ohki$^{3,}$\footnote{
E-mail address: ohki@kmi.nagoya-u.ac.jp} \ and \ 
Keigo~Sumita$^{1,}$\footnote{
E-mail address: k.sumita@moegi.waseda.jp
}\\*[20pt]
$^1${\it \normalsize 
Department of Physics, Waseda University, 
Tokyo 169-8555, Japan} \\
$^2${\it \normalsize 
Department of Physics, Kyoto University, 
Kyoto 606-8502, Japan} \\
$^3${\it \normalsize 
Kobayashi-Maskawa Institute for the 
Origin of Particles and the Universe (KMI),} \\
{\it \normalsize 
Nagoya University, Nagoya 464-8602, Japan} \\*[50pt]}

\date{
\centerline{\small \bf Abstract}
\begin{minipage}{0.9\linewidth}
\medskip 
\medskip 
\small
We present a four-dimensional (4D) ${\cal N}=1$ superfield 
description of supersymmetric Yang-Mills (SYM) theory in 
ten-dimensional (10D) spacetime with certain magnetic fluxes in 
compactified extra dimensions preserving partial ${\cal N}=1$ 
supersymmetry out of full ${\cal N}=4$. We derive a 4D effective 
action in ${\cal N}=1$ superspace directly from the 10D superfield 
action via dimensional reduction, and identify its dependence on 
dilaton and geometric moduli superfields. A concrete model for 
three generations of quark and lepton superfields are also shown. 
Our formulation would be useful for building various phenomenological 
models based on magnetized SYM theories or D-branes. 
\end{minipage}
}

\begin{titlepage}
\maketitle
\thispagestyle{empty}
\clearpage
\tableofcontents
\thispagestyle{empty}
\end{titlepage}

\renewcommand{\thefootnote}{\arabic{footnote}}
\setcounter{footnote}{0}

\section{Introduction}
Supersymmetric Yang-Mills (SYM) theories in higher-dimensional 
spacetime are quite attractive from both theoretical and  
phenomenological viewpoints. Such theories in ten-dimensional 
(10D) spacetime arise as low energy effective theories of 
superstrings, which are considered as the most promising candidates 
for an ultimate unification theory of elementary particles 
including gravitational interactions. 
It is indicated that SYM theories in various spacetime 
dimensions appear from D-branes in superstring theories. 
Various phenomenological superstring models beyond the standard 
model of elementary particles have been studied so far 
(for a review, see \cite{Ibanez}), 
many of which are based on SYM effective theories. 
Even without mentioning superstring theories, SYM theories in 
higher-dimensional spacetime themselves are interesting enough 
for phenomenological model buildings beyond the standard model. 

The standard model in four-dimensional (4D) spacetime may be 
realized at a low energy by compactifying the extra dimensions. 
In such a case, how to break higher-dimensional supersymmetry 
and to realize a chiral spectrum is the central issue. 
Toroidal compactifications with certain orbifold projections 
are mostly studied. Orbifolds may be considered as some singular 
limit of Calabi-Yau (CY) manifolds where most field contents, 
their wavefunctions and couplings in 4D effective theories 
are determined by geometric data. However, in general, such 
data are complicated and it is difficult to determine the 
CY metric explicitly, that makes phenomenological analyses 
on CY manifolds qualitative but not quantitative. 

Chiral spectra can be obtained not only on such nontrivial 
geometric backgrounds, but also even on a simple toroidal 
background with some gauge fluxes in Yang-Mills (YM) sector. 
(see Ref.~\cite{Cremades:2004wa} and references therein).\footnote{
 Other geometrical backgrounds with magnetic fluxes were also 
studied (see e.g. \cite{Conlon:2008qi}).}
In the latter case, chiral field contents and couplings are 
determined by the fluxes. For example, Yukawa couplings 
are determined by overlap integrals of the matter wavefunctions 
whose analytic forms are determined as functions of 
fluxes~\cite{Cremades:2004wa}. It is remarkable that the 
numbers of generations for chiral matter fields are determined 
by the fluxes, which provides a possibility that the flavor 
structure of the standard model is essentially determined 
by the YM-fluxes with sub-leading corrections from the 
background geometry. The observed flavor structures of 
quark and lepton masses and mixings would be realized by 
wavefunction localizations in extra dimensions due to the 
presence of magnetic fluxes~\cite{Cremades:2004wa,Abe:2008fi,Choi:2009pv} 
yielding some discrete flavor symmetries~\cite{Abe:2009vi}.\footnote{
Similar non-Abelian discrete flavor symmetries are obtained 
in heterotic orbifold models \cite{Kobayashi:2004ya}.} 
It is also remarkable that a certain class of magnetized 
D-branes can be regarded as the T-dual of some intersecting 
D-branes~\cite{Cremades:2004wa,Blumenhagen:2000wh}.

Here we focus on such YM-flux backgrounds which preserve 
(at least) 4D ${\cal N}=1$ supersymmetry out of the full 
higher-dimensional supersymmetry such as ${\cal N}=4$. 
In this case, field fluctuations around the background form 
${\cal N}=1$ supermultiplets and their action is written by 
superfields in ${\cal N}=1$ superspace, as indicated in 
Refs.~\cite{Marcus:1983wb,ArkaniHamed:2001tb} in the case 
without fluxes. In this paper we present a superfield 
description of SYM theories in higher-dimensional (especially 
10D) spacetime with magnetic fluxes in extra dimensions 
preserving 4D ${\cal N}=1$ supersymmetry, and derive 4D 
effective action for zero-modes in the superspace. 
Such a superfield description allows us to construct 
phenomenological models systematically and to analyze 
detailed low-energy properties of them, such as particle 
and superparticle flavor structures, features of Higgs 
particles and dark matter candidates, which are required 
for verifying these models by recent and upcoming data 
obtained from high-energy experiments as well as 
cosmological observations. 

The sections are organized as follows. 
In Sec.~\ref{sec:10d}, we rewrite the 10D SYM action in 4D 
${\cal N}=1$ superspace in the case with magnetic fluxes in 
extra dimensions, and derive zero-mode equations for superfields. 
In Sec.~\ref{sec:4d}, the 4D effective action for zero-modes 
are shown after a dimensional reduction. The action is extended 
to the case with local supersymmetry (i.e. supergravity) 
in Sec.~\ref{sec:sugra} where its moduli dependence is 
also identified. A direction of model building based on 
the effective action is indicated in Sec.~\ref{sec:model}. 
Finally, Sec.~\ref{sec:conc} is devoted to conclusions.

\section{10D magnetized SYM in ${\cal N}=1$ superspace}
\label{sec:10d}

In this section, we briefly review a compactification of 
10D SYM theory on flat 4D Minkowski spacetime times a 
product of factorizable three tori $T^2 \times T^2 \times T^2$. 
Then we derive a superfield description suitable for 
such a compactification with certain magnetic fluxes 
preserving 4D ${\cal N}=1$ supersymmetry. 
The geometric (torus) parameter dependence is 
explicitly shown in this procedure, which is important 
to determine couplings between YM and moduli superfields 
in the 4D effective action for zero-modes derived 
in the next sections.

\subsection{Toroidal compactification of 10D SYM theory}
\label{ssec:10dsym}

The action of 10D SYM theory is given by 
\begin{eqnarray}
S &=& \int d^{10}X\,\sqrt{-G}\,
\frac{1}{g^2}\,{\rm Tr}\left[ 
-\frac{1}{4}F^{MN}F_{MN} 
+\frac{i}{2} \bar\lambda \Gamma^M D_M \lambda 
\right], 
\label{eq:10dsym}
\end{eqnarray}
where $g$ is a 10D YM gauge coupling constant 
and the trace runs over the adjoint representation 
of a gauge group. The 10D spacetime coordinates are 
expressed by $X^M$, and the vector/tensor indices 
$M,N=0,1,\ldots,9$ are lowered and raised by the 10D 
metric $G_{MN}$ and its inverse $G^{MN}$, respectively. 
The YM field strength $F_{MN}$ and 
the covariant derivative $D_M$ are defined by 
\begin{eqnarray}
F_{MN} &=& 
\partial_M A_N - \partial_N A_M -i [A_M,A_N], 
\nonumber \\
D_M \lambda &=& \partial_M \lambda -i [A_M,\lambda], 
\nonumber
\end{eqnarray}
for a 10D vector (gauge) field $A_M$ and 
a 10D Majorana-Weyl spinor field $\lambda$ 
satisfying 
$\lambda^C = \lambda$ and 
$\Gamma \lambda = +\lambda$ 
where 
$\lambda^C$ is a 10D charge conjugation of $\lambda$, 
and $\Gamma$ is a 10D chirality operator. 

We decompose the 10D real coordinates $X^M=(x^\mu,y^m)$ into 
4D Minkowski spacetime coordinates $x^\mu$ with $\mu=0,1,2,3$ 
and six dimensional (6D) extra space coordinates $y^m$ with
$m=4,\ldots,9$.  
Note that $\mu=0$ describes the time component. Similarly 
the 10D vector field is decomposed as $A_M=(A_\mu,A_m)$. 
Our convention of the background metric is chosen as 
\begin{eqnarray}
ds^2 &=& G_{NN}dX^MdX^N \ = \ 
\eta_{\mu \nu} dx^\mu dx^\nu + g_{mn} dy^m dy^n, 
\nonumber
\end{eqnarray}
where 
$\eta_{\mu \nu}={\rm diag}(-1,+1,+1,+1)$. 
Then, we consider a torus compactification of internal 
6D space $y^m$ by identifying $y^m \sim y^m + 2$. 
The 6D torus is further assumed as a product of 
factorizable three tori, $T^2 \times T^2 \times T^2$, 
and then the extra 6D metric can be described as 
\begin{eqnarray}
g_{mn} &=& 
\begin{pmatrix}
g^{(1)} & 0 & 0 \\
0 & g^{(2)}& 0 \\
0 & 0 & g^{(3)} 
\end{pmatrix}, 
\nonumber
\end{eqnarray}
where each of entries is a $2 \times 2$ matrix and 
submatrices in the diagonal part are given by 
\begin{eqnarray}
g^{(i)} &=& (2 \pi R_i)^2 
\begin{pmatrix}
1 & {\rm Re}\,\tau_i \\
{\rm Re}\,\tau_i & |\tau_i|^2
\end{pmatrix}, 
\label{eq:gtorus}
\end{eqnarray}
for $i=1,2,3$. 
The real and complex parameters $R_i$ and $\tau_i$, 
respectively, determine the size and the shape of $i$th 
torus $T^2$. Especially the area ${\cal A}^{(i)}$ of the 
$i$th torus is described as 
\begin{eqnarray}
{\cal A}^{(i)} &=& (2 \pi R_i)^2\,{\rm Im}\, \tau_i. 
\nonumber
\end{eqnarray}

For convenience in a supersymmetric world, we introduce 
complex (extra-dimensional) coordinates $z^i$ for $i=1,2,3$ 
and also complex vector components $A_i$ defined by 
\begin{eqnarray}
z^i &\equiv& \frac{1}{2}(y^{2+2i} + \tau_i\,y^{3+2i}), 
\qquad 
\bar{z}^{\bar{i}} \ \equiv \ (z^i)^\ast, 
\label{eq:cczi} \\
A_i &\equiv& 
-\frac{1}{{\rm Im}\,\tau_i} 
(\tau_i^\ast \,A_{2+2i}-A_{3+2i}), 
\qquad 
\bar{A}_{\bar{i}} \ \equiv \ (A_i)^\dagger. 
\nonumber
\end{eqnarray}
The torus boundary conditions are given by 
$z^i \sim z^i +1$ and $z^i \sim  z^i +\tau^i$. 
A metric $h_{i\bar{j}}$ for the complex coordinates 
is extracted from 
\begin{eqnarray}
ds_{6D}^2 &=& g_{mn}dy^mdy^n 
\ \equiv \ 2h_{i\bar{j}}dz^id\bar{z}^{\bar{j}}, 
\nonumber
\end{eqnarray}
and then we find 
\begin{eqnarray}
h_{i\bar{j}} &=& 
2\,(2 \pi R_i)^2\,\delta_{i\bar{j}} 
\ = \ \delta_{{\rm i}\bar{\rm j}}\,
e_i^{\ {\rm i}}\,\bar{e}_{\bar{j}}^{\ \bar{\rm j}}, 
\nonumber
\end{eqnarray}
where $e_i^{\ {\rm i}}$ is a vielbein defined by 
\begin{eqnarray}
e_i^{\ {\rm i}} &=& 
\sqrt{2}\,(2 \pi R_i)\,\delta_i^{\ {\rm i}}, 
\nonumber
\end{eqnarray}
with its inverse $e_{{\rm i}}^{\ i}$ 
and complex conjugate 
$\bar{e}_{\bar{i}}^{\ \bar{\rm i}}$ 
satisfying 
\begin{eqnarray}
e_{{\rm i}}^{\ i} e_i^{\ {\rm j}} 
&=& \delta_{{\rm i}}^{\ {\rm j}}, \qquad 
e_i^{\ {\rm i}} e_{{\rm i}}^{\ j} 
\ = \ \delta_i^{\ j}, \qquad 
\bar{e}_{\bar{i}}^{\ \bar{\rm i}} 
\ = \ (e_i^{\ {\rm i}})^\ast, 
\nonumber
\end{eqnarray}
and Roman indices representing local Lorentz space. 
The Italic (Roman) indices $i,j,\ldots$ 
(${\rm i}$, ${\rm j}, \ldots$) are lowered and raised 
by the metric $h_{i\bar{j}}$ 
and its inverse $h^{\bar{i}j}$ 
($\delta_{{\rm i}\bar{\rm j}}$ 
and its inverse $\delta^{\bar{\rm i}{\rm j}}$), 
respectively.

\subsection{Superfield description on nontrivial gauge background}
\label{ssec:sfss}

The 10D SYM theory has ${\cal N}=4$ supersymmetry in terms 
of a 4D supercharge. The YM fields $A_M$ and $\lambda$ can 
be decomposed into 4D ${\cal N}=1$ (on-shell) supermultiplets as 
\begin{eqnarray}
{\bm V} &=& \left\{ A_\mu, \lambda_0 \right\}, \qquad 
{\bm \phi}_i \ = \ \left\{ A_i, \lambda_i \right\}. 
\nonumber
\end{eqnarray}
Here the 10D Majorana-Weyl spinor $\lambda$ is decomposed 
into four 4D Weyl (or equivalently Majorana) spinors 
$\lambda_0$ and $\lambda_i$ satisfying 
\begin{eqnarray}
\Gamma^{(i)} \lambda_0 &=& +\lambda_0, \qquad 
\Gamma^{(i)} \lambda_j \ = \ +\lambda_j \quad (i=j), \qquad 
\Gamma^{(i)} \lambda_j \ = \ -\lambda_j \quad (i \ne j), 
\label{eq:gamma5i}
\end{eqnarray}
where $\Gamma^{(i)}$ for each $i=1,2,3$ are a chirality 
operator associated with 6D spacetime coordinates $(x^\mu,z^i)$. 
Note that an eigenvalue of the 10D chirality operator $\Gamma$ 
is obtained by a product of the eigenvalues of 
$\Gamma^{(1)}$, $\Gamma^{(2)}$ and $\Gamma^{(3)}$. 
If we write the chirality of $i$th complex coordinate $z_i$ 
in the $i$th subscript of $\lambda$ like 
$\lambda_{\pm \pm \pm}$, 
the decomposed spinor fields $\lambda_0$ and $\lambda_i$ 
are identified with the chirality eigenstate 
$\lambda_{\pm \pm \pm}$ as 
\begin{eqnarray}
\lambda_0 &=& \lambda_{+++}, \qquad 
\lambda_1 \ = \ \lambda_{+--}, \qquad 
\lambda_2 \ = \ \lambda_{-+-}, \qquad 
\lambda_3 \ = \ \lambda_{--+}. 
\nonumber
\end{eqnarray}
Note that the components 
$\lambda_{---}$, $\lambda_{-++}$, 
$\lambda_{+-+}$ and $\lambda_{++-}$ 
do not exist in the 10D Majorana-Weyl spinor $\lambda$ 
due to the condition $\Gamma \lambda = +\lambda$. 

The above ${\cal N}=1$ vector multiplet ${\bm V}$ and 
chiral multiplets ${\bm \phi}_i$ are expressed, respectively, 
by a vector superfield $V$ and a chiral superfield $\phi_i$. 
Our definition of superfields is as follows: 
\begin{eqnarray}
V &\equiv& -\theta\sigma^\mu\bar\theta A_\mu
+i\bar\theta\bar\theta\theta\lambda_0 
-i\theta\theta\bar\theta\bar\lambda_0 
+\frac12\theta\theta\bar\theta\bar\theta D, 
\nonumber \\
\phi_i &\equiv& \frac1{\sqrt2} A_i
+\sqrt2\theta\lambda_i+\theta\theta F_i, 
\nonumber
\end{eqnarray}
where $\theta$ and $\bar\theta$ are 
Grassmann coordinates of 4D ${\cal N}=1$ superspace with 
spinor indices $\alpha$ and $\dot\alpha$ omitted 
respectively\footnote{Our conventions mostly 
follow those in Ref.~\cite{Wess:1992cp}.}. 
The superfield description allows us to express the 
10D SYM action (\ref{eq:10dsym}) in ${\cal N}=1$ 
superspace as~\cite{ArkaniHamed:2001tb} 
\begin{eqnarray}
S &=& \int d^{10}X \sqrt{-G} \left[ \int d^4\theta\,{\cal K}
+\left\{ \int d^2 \theta\,\left( 
\frac{1}{4g^2} {\cal W}^\alpha {\cal W}_\alpha 
+ {\cal W} \right) +{\rm h.c.} \right\} 
\right], 
\label{eq:10dss}
\end{eqnarray}
where 
\begin{eqnarray}
{\cal K} &=& \frac{2}{g^2}  h^{\bar{i}j} 
{\rm Tr} \left[\left(\sqrt2\bar\partial_{\bar{i}}
+\bar\phi_{\bar{i}}\right)e^{-V}
\left(-\sqrt2\partial_j + \phi_j \right)e^V
+\bar\partial_{\bar{i}} e^{-V} \partial_j e^V \right] 
+{\cal K}_{\rm WZW}, 
\nonumber \\
{\cal W} &=& \frac{1}{g^2} 
\epsilon^{{\rm i}{\rm j}{\rm k}} 
e_{{\rm i}}^{\ i} e_{{\rm j}}^{\ j} e_{{\rm k}}^{\ k} 
{\rm Tr} \left[ \sqrt{2}\, \phi_i
\left(\partial_j\phi_k-\frac{1}{3\sqrt2}
\left[\phi_j,\phi_k\right]\right)\right], 
\nonumber
\end{eqnarray}
with a totally antisymmetric tensor 
$\epsilon^{{\rm i}{\rm j}{\rm k}}$ 
and $\epsilon^{123}=1$. 
The field strength superfield ${\cal W}_\alpha$ is defined by 
${\cal W}_\alpha \equiv 
-\frac{1}{4} \bar{D} \bar{D} e^{-V} D_\alpha e^V$ 
where $D_\alpha$ and $\bar{D}_{\dot\alpha}$ 
are a supercovariant derivative and its conjugate with 
4D spinor indices $\alpha$ and $\dot\alpha$, respectively. 
The term ${\cal K}_{\rm WZW}$ corresponds to a 
Wess-Zumino-Witten term 
which vanishes in Wess-Zumino (WZ) gauge. 
The equations of motion for 
auxiliary fields $D$ and $F_i$ lead to 
\begin{eqnarray}
D &=& - h^{\bar{i}j} \left( 
\bar\partial_{\bar{i}} A_j + \partial_j \bar{A}_{\bar{i}} 
+\frac{1}{2} \left[ \bar{A}_{\bar{i}}, A_j \right] \right), 
\label{eq:osd} \\
\bar{F}_{\bar{i}} &=& -h_{j\bar{i}}\, 
\epsilon^{{\rm j}{\rm k}{\rm l}} 
e_{{\rm j}}^{\ j} e_{{\rm k}}^{\ k} e_{{\rm l}}^{\ l} 
\left( \partial_k A_l 
-\frac{1}{4} \left[ A_k,\, A_l \right] \right). 
\label{eq:osfi}
\end{eqnarray}
The original SYM action (\ref{eq:10dsym}) is obtained 
after integrating over the superspace in Eq.~(\ref{eq:10dss}) 
and substituting on-shell values 
(\ref{eq:osd}) and (\ref{eq:osfi}). 

We assume 4D Lorentz invariant and (at least ${\cal N}=1$) 
supersymmetric VEVs of fields, 
\begin{eqnarray}
\vev{A_i} &\ne& 0, \qquad 
\vev{A_\mu} \ = \ 
\vev{\lambda_0} \ = \ 
\vev{\lambda_i} \ = \ 
\vev{F_i} \ = \ 
\vev{D} \ = \ 0. 
\label{eq:vevs}
\end{eqnarray}
We will see later\footnote{See 
Eqs.(\ref{eq:abd}), (\ref{eq:abfi}) and (\ref{eq:susy}).} 
that certain magnetized background satisfy Eq.~(\ref{eq:vevs}). 
Then we extract fluctuations $\tilde{V}$ and $\tilde{\phi}_i$ 
of superfields $V$ and $\phi_i$, respectively, around the 
vacuum configuration (\ref{eq:vevs}) as 
\begin{eqnarray}
V &\equiv& \vev{V} + \tilde V, \qquad 
\phi_i \ \equiv \ \vev{\phi_i} + \tilde\phi_i, 
\nonumber
\end{eqnarray}
where $\vev{V}=0$ and $\vev{\phi_i}=\vev{A_i}/\sqrt{2}$ 
due to Eq.~(\ref{eq:vevs}). 
For a notational convenience, in the following, 
we omit tildes and use original notation $V$ and $\phi_i$ 
for their corresponding fluctuations around the vacuum. 
Then, in the WZ-gauge, the functions 
${\cal K}$ and ${\cal W}$ are expanded in powers of $V$ as 
\begin{eqnarray}
{\cal K} &=& \frac{2}{g^2} h^{\bar{i}j} {\rm Tr} \Bigg[ 
\bar\phi_{\bar{i}} \phi_j 
+\sqrt{2} \left\{ \left(\bar\partial_{\bar{i}} \phi_j
+\frac{1}{\sqrt2} 
\left[ \vev{\bar\phi_{\bar{i}}},\,\phi_j \right] 
+{\rm h.c.} \right)
+\frac{1}{\sqrt2} 
\left[ \bar\phi_{\bar{i}},\,\phi_j \right] \right\} V 
\nonumber \\ && \qquad \qquad 
+(\bar\partial_{\bar{i}}V)(\partial_j V) 
+ \frac{1}{2} \left( \bar\phi_{\bar{i}} \phi_j 
+\phi_j \bar\phi_{\bar{i}} \right) V^2 
-\bar\phi_{\bar{i}} V \phi_j V \Bigg] 
+{\cal K}^{({\rm D})}
+{\cal K}^{({\rm br})}, 
\label{eq:kfv} \\
{\cal W} &=& \frac{1}{g^2} \epsilon^{{\rm i}{\rm j}{\rm k}} 
e_{{\rm i}}^{\ i} e_{{\rm j}}^{\ j} e_{{\rm k}}^{\ k}
{\rm Tr} \left[ \sqrt{2} \left( 
\partial_i\phi_j -\frac{1}{\sqrt{2}} 
\left[ \vev{\phi_i},\,\phi_j \right] \right) \phi_k 
-\frac{2}{3} \phi_i \phi_j \phi_k \right]+{\cal W}^{(\rm{F})}, 
\label{eq:wfv}
\end{eqnarray}
where 
\begin{eqnarray}
{\cal K}^{({\rm br})} &=& 
\frac{1}{g^2} h^{\bar{i}j} {\rm Tr} \Bigg[ 
\frac{1}{2} \left( 
\vev{\bar\phi_{\bar{i}}} \vev{\phi_j} 
+\vev{\phi_j} \vev{\bar\phi_{\bar{i}}} 
\vev{\bar\phi_{\bar{i}}} \phi_j 
+\phi_j \vev{\bar\phi_{\bar{i}}} 
+\bar\phi_{\bar{i}} \vev{\phi_j} 
+\vev{\phi_j} \bar\phi_{\bar{i}} 
\right) V^2 
\nonumber \\ && \qquad \qquad 
-\vev{\bar\phi_{\bar{i}}} V \vev{\phi_j} V
-\vev{\bar\phi_{\bar{i}}} V \phi_j V
-\bar\phi_{\bar{i}} V \vev{\phi_j} V
 \Bigg], 
\label{eq:unb}
\end{eqnarray}
and we omit terms in higher-powers of 
$\theta$ and $\bar\theta$ than 
$\theta^2 \bar\theta^2$ which vanish 
in the superspace action (\ref{eq:10dss}). 
The two terms of ${\cal K}^{({\rm D})}$ and ${\cal W}^{(\rm{F})}$  
are eliminated in the case that 
the SUSY conditions (\ref{eq:osd}) and (\ref{eq:osfi}) are satisfied.   
The vacuum configuration (\ref{eq:vevs}) in general 
breaks the gauge symmetry (as well as the 
higher-dimensional supersymmetry) of original SYM theory 
down to its subgroup, and vector multiplets associated 
with the broken symmetries become massive. Such masses 
are generated in ${\cal K}^{({\rm br})}$.

\subsection{Zero-mode equations in superspace}
\label{ssec:zme}

On the ${\cal N}=1$ supersymmetric background (\ref{eq:vevs}), 
Kaluza-Klein (KK) mode-equations should be written as superfield 
equations. Then, mode expansions of superfields $V$ and $\phi_j$ 
are expressed as 
\begin{eqnarray}
V(x^\mu,{\bm z},\bar{\bm z}) 
&=& \sum_{{\bm n}} \left( \prod_i  
f_0^{(i),n_i}(z^i,\bar{z}^{\bar{i}}) \right) 
V^{{\bm n}}(x^\mu), 
\label{eq:mev} \\
\phi_j(x^\mu,{\bm z},\bar{\bm z}) 
&=& \sum_{{\bm n}} \left( \prod_i  
f_j^{(i),n_i}(z^i,\bar{z}^{\bar{i}}) \right) 
\phi_j^{{\bm n}}(x^\mu), 
\label{eq:mephi}
\end{eqnarray}
where 
${\bm z}=(z^1,z^2,z^3)$, 
$\bar{\bm z}=(
\bar{z}^{\bar{1}},\bar{z}^{\bar{2}},\bar{z}^{\bar{3}})$, 
${\bm n} = (n_1, n_2, n_3)$ 
and $n_i \in {\bf Z}$ for $i,j=1,2,3$. 
The 4D fields $V^{{\bm n}}$ and $\phi_j^{{\bm n}}$ 
are KK modes of 10D fields $V$ and $\phi_j$, respectively, 
with KK momenta of three tori labeled by ${\bm n}$. 
The functions $f_0^{(i),n_i}$ and $f_j^{(i),n_i}$ are 
wavefunctions of $V^{{\bm n}}$ and $\phi_j^{{\bm n}}$, 
respectively, in the $i$th torus. 
Note that all the quantities 
$\{V, \phi_j, V^{{\bm n}}, \phi_j^{{\bm n}}, 
f_0^{(i),n_i}, f_j^{(i),n_i} \}={\cal G}$ 
in Eqs.~(\ref{eq:mev}) and (\ref{eq:mephi}) 
have the same subscripts of YM indices implicitly, 
e.g., 
${\cal G}={\cal G}_{AB}$ with $A,B=1,2,\ldots,N$ 
for $U(N)$ SYM theory, 
because the both-hand sides of 
Eqs.~(\ref{eq:mev}) and (\ref{eq:mephi}) 
are adjoint matrices of YM gauge group, 
and wavefunctions can be different from each other 
element by element due to a possible gauge symmetry 
breaking caused by $\vev{A_i} \ne 0$ in Eq.~(\ref{eq:vevs}).  
We should remark that $V^{{\bm n}}$ and $\phi_j^{{\bm n}}$ 
are a vector and chiral superfields, respectively, while 
$f_0^{(i),n_i}$ and $f_j^{(i),n_i}$ are independent of  
superspace coordinates $\theta$ and $\bar\theta$ on a 
${\cal N}=1$ supersymmetric background (\ref{eq:vevs}). 

Substituting Eqs.~(\ref{eq:mev}) and (\ref{eq:mephi}) 
into the superspace action (\ref{eq:10dss}) 
with Eqs.~(\ref{eq:kfv}) and (\ref{eq:wfv}), 
we find zero-mode equations as 
\begin{eqnarray}
\partial_i f_0^{(i),n_i=0} -\frac1{\sqrt2}
\left[ \vev{\phi_i},\,f_0^{(i),n_i=0} \right]
&=& 0, 
\nonumber \\
\bar\partial_{\bar{i}} f_j^{(i),n_i=0} +\frac1{\sqrt2} 
\left[ \vev{\bar\phi_{\bar{i}}},\,f_j^{(i),n_i=0} \right] 
&=& 0 \qquad (i=j), 
\label{eq:kkeii} \\
\partial_i f_j^{(i),n_i=0} -\frac1{\sqrt2}
\left[ \vev{\phi_i},\,f_j^{(i),n_i=0} \right]
&=& 0 \qquad (i \ne j), 
\label{eq:kkeij}
\end{eqnarray}
with which terms quadratic in superfields $\phi_i$ and $V$ 
are eliminated in Eqs.~(\ref{eq:kfv}) and (\ref{eq:wfv}) 
except in Eq.~(\ref{eq:unb}), and then vanishing masses
for zero-modes $\phi_j^{{\bm n}={\bm 0}}$ are guaranteed. 
Note that a sign-difference between the second terms in the 
left-hand sides of Eqs.~(\ref{eq:kkeii}) and (\ref{eq:kkeij}) 
comes from the chirality structure (\ref{eq:gamma5i}). 
A remarkable point is that zero-mode equations are written 
for superfields. Such a superfield description allows us 
to perform a dimensional reduction while keeping a 
manifest ${\cal N}=1$ superspace structure preserved by 
the gauge background (\ref{eq:vevs}).

\section{4D effective action}
\label{sec:4d}

In the following, we focus on 4D massless modes (called zero-modes) 
in chiral superfields $\phi_i$ which possess nontrivial wavefunction 
profiles under the existence of magnetic fluxes in extra dimensions. 
We derive a 4D effective action for zero-modes by solving the zero-mode 
equations and performing integrations over extra space coordinates 
in the 10D action, while keeping the ${\cal N}=1$ superspace structure 
preserved by the flux background. 

\subsection{Dimensional reduction with magnetic fluxes}
\label{ssec:dimred}

Because we focus on zero-modes in the following, 
for a notational convenience, their wavefunctions are denoted as 
\begin{eqnarray}
f_0^{(i),n_i=0} &\equiv& f_0^{(i)}, \qquad 
f_j^{(i),n_i=0} \ \equiv \ f_j^{(i)}. 
\nonumber
\end{eqnarray}
After substituting $\vev{\phi_i}=\vev{A_i}/\sqrt{2}$ 
into Eqs.~(\ref{eq:kkeii}) and (\ref{eq:kkeij}), 
zero-mode equations are expressed by 
\begin{eqnarray}
\bar\partial_{\bar{i}} f_j^{(i)} +\frac{1}{2} 
\left[ \vev{\bar{A}_{\bar{i}}},\,f_j^{(i)} \right] 
&=& 0 \qquad (i=j), 
\label{eq:zmeii} \\
\partial_i f_j^{(i)} -\frac{1}{2}
\left[ \vev{A_i},\,f_j^{(i)} \right]
&=& 0 \qquad (i \ne j). 
\label{eq:zmeij}
\end{eqnarray}
It is found that Eqs.(\ref{eq:zmeii}) and (\ref{eq:zmeij}) 
are equivalent to zero-mode equations for charged fermion 
fields $\lambda_j$ in the gauge background~\cite{Cremades:2004wa} 
as it should be. 

In the following we consider the case that the YM gauge group is 
$U(N)$ and assume a supersymmetric (Abelian) magnetic background: 
\begin{eqnarray}
\vev{A_i} &=& \frac{\pi}{{\rm Im}\, \tau_i} 
\left( M^{(i)}\,\bar{z}_{\bar{i}}+\bar\zeta_i \right), 
\nonumber
\end{eqnarray}
where 
\begin{eqnarray}
M^{(i)} &=& {\rm diag}
(M_1^{(i)},M_2^{(i)},\ldots,M_N^{(i)}), 
\nonumber \\
\zeta_i &=& {\rm diag}
(\zeta_1^{(i)},\zeta_2^{(i)},\ldots,\zeta_N^{(i)}), 
\nonumber
\end{eqnarray}
are $N \times N$ diagonal matrices of Abelian 
magnetic fluxes and Wilson-lines, respectively.
Here, the magnetic fluxes must satisfy  
$M_1^{(i)},M_2^{(i)},\ldots,M_N^{(i)} \in {\bf Z}$ 
due to the Dirac's quantization condition. 
We assume that the above magnetic fluxes 
are further constrained in such a way 
that they satisfy supersymmetry condition 
\begin{eqnarray}
h^{\bar{i}j} \left( \bar\partial_{\bar{i}} \vev{A_j} 
+ \partial_j \vev{\bar{A}_{\bar{i}}} \right) &=& 0, 
\label{eq:abd} \\
\epsilon^{{\rm j}{\rm k}{\rm l}} 
e_{{\rm k}}^{\ k} e_{{\rm l}}^{\ l} 
\partial_k \vev{A_l} &=& 0, 
\label{eq:abfi}
\end{eqnarray}
in order to satisfy Eq.~(\ref{eq:vevs}) with 
Eqs.~(\ref{eq:osd}) and (\ref{eq:osfi}) 
for the Abelian background. 

In the case that all the magnetic fluxes 
$M_1^{(i)},M_2^{(i)},\ldots,M_N^{(i)}$ take different values 
from each other, the YM gauge symmetry is broken down as 
$U(N) \to U(1)^N$ by the existence of fluxes. 
On the other hand, if any of them degenerate, e.g., 
\begin{eqnarray}
&M_1^{(i)}=M_2^{(i)}=\cdots=M_{N_1}^{(i)},& 
\nonumber \\
&M_{N_1+1}^{(i)}=M_{N_1+2}^{(i)}=\cdots=M_{N_1+N_2}^{(i)},& 
\nonumber \\
&\vdots& 
\nonumber \\
&M_{N_1+N_2+\cdots+N_{\tilde{N}-1}+1}^{(i)}
=M_{N_1+N_2+\cdots+N_{\tilde{N}-1}+2}^{(i)}=\cdots
=M_{N_1+N_2+\cdots+N_{\tilde{N}-1}+N_{\tilde{N}}}^{(i)},& 
\label{eq:gsb}
\end{eqnarray}
with $\sum_a N_a=N$ and $M_{N_a} \ne M_{N_b}$ 
for $a,b=1,2,\ldots,\tilde{N}$ and $a \ne b$, 
the breaking pattern is changed as 
$U(N) \to \prod_a U(N_a)$. 
The same holds for Wilson-lines. 
In the following, indices $a,b,c=1,2,\ldots,\tilde{N}$ label the 
unbroken YM subgroups on the flux and Wilson-line background, 
and traces in expressions are performed within such subgroups. 

{}From Eqs.~
(\ref{eq:zmeii}) and (\ref{eq:zmeij}), 
for the unbroken YM subgroup labeled by $a$ and $b$, 
we find zero-mode equations 
$(f_j^{(i)})_{ab}$ as 
\begin{eqnarray}
\left[ \bar\partial_{\bar{i}} +\frac{\pi}{2{\rm Im}\,\tau_i} 
\left( M^{(i)}_{ab}\,z_i 
+\zeta^{(i)}_{ab} \right) \right] (f_j^{(i)})_{ab} 
&=& 0 \qquad (i=j), 
\label{eq:czmeii} \\
\left[ \partial_i -\frac{\pi}{2{\rm Im}\,\tau_i}
\left( M^{(i)}_{ab}\,\bar{z}_{\bar{i}} 
+\bar\zeta^{(i)}_{ab} \right) \right] (f_j^{(i)})_{ab} 
&=& 0 \qquad (i \ne j), 
\label{eq:czmeij}
\end{eqnarray}
where 
\begin{eqnarray}
M^{(i)}_{ab} &=& M^{(i)}_{N_a}-M^{(i)}_{N_b}, \qquad 
\zeta^{(i)}_{ab}=\zeta^{(i)}_{N_a}-\zeta^{(i)}_{N_b}. 
\nonumber
\end{eqnarray}

For $i=j$, a normalizable solution of 
Eq.~(\ref{eq:czmeii}) is found as 
\begin{eqnarray}
(f_j^{(i)})_{ab} &=& 
f^{I^{(i)}_{ab}} \ \equiv \ 
\left\{ \begin{array}{ll}
\displaystyle 
\Theta^{I^{(i)}_{ab},\,M^{(i)}_{ab}} 
\left( \tilde{z}_i \right) 
& \quad (M^{(i)}_{ab} > 0) \\
\displaystyle ({\cal A}^{(i)})^{-1/2} & 
\quad (M^{(i)}_{ab} = 0) \\
\displaystyle 
0 & \quad (M^{(i)}_{ab} < 0) 
\end{array} \right., 
\label{eq:fijab}
\end{eqnarray}
where 
$\tilde{z}_i \equiv 
z_i + \frac{\zeta^{(i)}_{ab}}{M^{(i)}_{ab}}$ and 
\begin{eqnarray}
I^{(i)}_{ab} &=& 
\left\{ \begin{array}{ll}
1,\, \ldots,\, \left| M^{(i)}_{ab} \right| 
& \quad (M^{(i)}_{ab} > 0) \\
0 & \quad (M^{(i)}_{ab} = 0) \\
{\rm no \ solution} & \quad (M^{(i)}_{ab} < 0) 
\end{array} \right.. 
\label{eq:igen}
\end{eqnarray}
Due to the effect of chirality projection by fluxes, 
no zero-mode appears for $M^{(i)}_{ab} < 0$. 
On the other hand, for  $M^{(i)}_{ab} > 0$, there 
appear $M^{(i)}_{ab}$ zero-modes, and these zero-modes are 
labeled by the index $I^{(i)}_{ab}$.
In Eq.~(\ref{eq:fijab}), the wavefunction profile 
$\Theta^{I,\,M}(z)$ is determined as 
\begin{eqnarray}
\Theta^{I,\,M}(z) &\equiv& 
{\cal N}_M \, e^{\pi i 
\frac{{\rm Im}\,z}{{\rm Im}\,\tau} M z}\,
\vartheta 
\begin{bmatrix} \frac{I}{M} \\ 0 \end{bmatrix} 
\left( M z, M \tau \right), 
\nonumber
\end{eqnarray}
where $\vartheta$ represents the 
Jacobi theta-function: 
\begin{eqnarray}
\vartheta 
\begin{bmatrix} a \\ b \end{bmatrix} 
\left( \nu, \tau \right) &=& 
\sum_{l \in {\bf Z}} 
e^{\pi i \left( a + l \right)^2 \tau}
e^{2 \pi i \left( a + l \right) 
\left( \nu + b \right)} .
\nonumber
\end{eqnarray}
The normalization constant ${\cal N}_M$ is found as 
\begin{eqnarray}
{\cal N}_{M^{(i)}} &=& \left( 
\frac{2\,{\rm Im}\,\tau_i\,|M^{(i)}|}{({\cal A}^{(i)})^2} 
\right)^{1/4}, 
\nonumber
\end{eqnarray}
from a normalization condition, 
\begin{eqnarray}
\int dz_i d \bar{z}_i \sqrt{{\rm det}\,g^{(i)}}\,
f^I\,(f^J)^\ast &=& \delta^{IJ}, 
\nonumber
\end{eqnarray}
for $I,J \ne 0$ or $I=J=0$. 

For $i \ne j$, on the other hand, 
from Eq.~(\ref{eq:czmeij}) we find 
\begin{eqnarray}
(f_j^{(i)})_{ab} &=& 
f^{I^{(i)}_{ab}} \ \equiv \ 
\left\{ \begin{array}{ll}
\displaystyle 
0 & \quad (M^{(i)}_{ab} > 0) \\
\displaystyle ({\cal A}^{(i)})^{-1/2} & 
\quad (M^{(i)}_{ab} = 0) \\
\displaystyle 
\left( \Theta^{I^{(i)}_{ab},\,|M^{(i)}_{ab}|} 
\left( \tilde{z}_i \right) \right)^\ast 
& \quad (M^{(i)}_{ab} < 0) 
\end{array} \right., 
\nonumber
\end{eqnarray}
where 
$\tilde{z}_i \equiv 
z_i - \frac{\zeta^{(i)}_{ab}}{|M^{(i)}_{ab}|}$, 
instead of Eq.~(\ref{eq:fijab}).

\subsection{4D effective action for zero-modes}
\label{ssec:zm4deff}

Due to the gauge symmetry breaking 
$U(N) \to \prod_a U(N_a)$ with $a=1,\ldots,\tilde{N}$
caused by the fluxes satisfying Eq.~(\ref{eq:gsb}), 
off-diagonal elements $(V^{{\bm n}={\bm 0}})_{ab}$ 
($a \ne b$) obtain mass terms in Eq.~(\ref{eq:unb}), 
while diagonal elements 
$(V^{{\bm n}={\bm 0}})_{aa}$ do not. 
Then, we express the zero-modes 
$(V^{{\bm n}={\bm 0}})_{aa}$, 
which contain gauge fields for the unbroken gauge symmetry 
$\prod_a U(N_a)$, as 
\begin{eqnarray}
(V^{{\bm n}={\bm 0}})_{aa} &\equiv& V^a. 
\nonumber
\end{eqnarray}
On the other hand, 
from Eq.~(\ref{eq:igen}), 
for $^\exists j \ne i$ with 
$M^{(j)}_{ab} < 0$ and $M^{(i)}_{ab} > 0$ 
we find the zero-mode field $(\phi_j^{{\bm n}={\bm 0}})_{ab}$ 
degenerates with a total degeneracy 
$N_{ab}=\prod_i \left| M^{(i)}_{ab} \right|$, 
while $(\phi_j^{{\bm n}={\bm 0}})_{ba}$ has no zero-mode solution, 
yielding 4D chiral generations in the $ab$-sector. 
The opposite is true for 
$M^{(j)}_{ab} > 0$ and $M^{(i)}_{ab} < 0$ 
yielding 4D chiral generations in the $ba$-sector. 
Therefore, we denote the zero-modes 
$(\phi_j^{{\bm n}={\bm 0}})_{ab}$ 
with the degeneracy $N_{ab}$ as 
\begin{eqnarray}
(\phi_j^{{\bm n}={\bm 0}})_{ab} &\equiv& 
g\,\phi_j^{{\cal I}_{ab}}, 
\nonumber
\end{eqnarray}
where ${\cal I}_{ab} \equiv 
(I^{(1)}_{ab}, I^{(2)}_{ab}, I^{(3)}_{ab})$ labels 
the degeneracy (generations). 
We normalize $\phi_j^{{\cal I}_{ab}}$ by 
the 10D YM coupling constant $g$. 

The analytic expressions of zero-mode wavefunctions 
allow us to derive 4D effective action after substituting 
the mode expansion (\ref{eq:mephi})
and integrating 10D action (\ref{eq:10dss}) 
over 6D extra space coordinates $z_i$ and $\bar{z}^{\bar{i}}$. 
The effective action for zero-mode chiral superfields $\phi_j^{{\cal
I}_{ab}}$ is then found as 
\begin{eqnarray}
S_{\rm eff} &=& 
\int d^4x \left[ \int d^4 \theta\,{\cal K}_{\rm eff} 
+\left\{ \int d^2 \theta \left( 
\sum_a \frac{1}{4g_a^2} \,W^{a,\alpha} W^a_\alpha 
+{\cal W}_{\rm eff} \right) 
+{\rm h.c.} \right\} \right], 
\label{eq:seff}
\end{eqnarray}
where 
\begin{eqnarray}
{\cal K}_{\rm eff} &=& \sum_{i,j}\,\sum_{a,b}\,
\sum_{{\cal I}_{ab}} 
\tilde{Z}_{{\cal I}_{ab}}^{\bar{i}j} {\rm tr} \left[ 
\bar\phi_{\bar{i}}^{{\cal I}_{ab}} e^{-V^a} 
\phi_j^{{\cal I}_{ab}} e^{V^b} \right], 
\nonumber \\
{\cal W}_{\rm eff} &=& 
\sum_{i,j,k}\,\sum_{a,b,c}\,
\sum_{{\cal I}_{ab},\,
{\cal I}_{bc},\,
{\cal I}_{ca}}
\tilde\lambda_{{\cal I}_{ab}
{\cal I}_{bc}
{\cal I}_{ca}}^{ijk} 
{\rm tr} \left[ 
\phi_i^{{\cal I}_{ab}} 
\phi_j^{{\cal I}_{bc}} 
\phi_k^{{\cal I}_{ca}} \right], 
\nonumber
\end{eqnarray}
and 
\begin{eqnarray}
W_\alpha^a &=& -\frac{1}{4} \bar{D} \bar{D} 
e^{-V^a} D_\alpha e^{V^a}, \qquad 
g_a \ = \ g \left( \prod_i {\cal A}^{(i)} \right)^{-1/2}. 
\nonumber
\end{eqnarray}
The K\"ahler metric $\tilde{Z}_{{\cal I}_{ab}}$ 
and the holomorphic Yukawa couplings 
$\tilde\lambda_{{\cal I}_{ab}
{\cal I}_{bc}
{\cal I}_{ca}}$ 
can be written as 
\begin{eqnarray}
\tilde{Z}_{{\cal I}_{ab}}^{\bar{i}j} 
&=& 2 h^{\bar{i}j}, 
\label{eq:km} \\
\tilde\lambda_{
{\cal I}_{ab}
{\cal I}_{bc}
{\cal I}_{ca}}^{ijk} &=& 
-\frac{2g}{3} 
\epsilon^{{\rm i}{\rm j}{\rm k}} 
e_{{\rm i}}^{\ i} 
e_{{\rm j}}^{\ j} 
e_{{\rm k}}^{\ k} 
\prod_r 
\tilde\lambda^{(r)}_{
I^{(r)}_{ab}I^{(r)}_{bc}I^{(r)}_{ca}}, 
\label{eq:yc}
\end{eqnarray}
where $r=1,2,3$ and 
\begin{eqnarray}
\tilde\lambda^{(r)}_{
I^{(r)}_{ab}I^{(r)}_{bc}I^{(r)}_{ca}} 
&=& \int d^2z^r \sqrt{{\rm det}\,g^{(r)}}\, 
f^{I^{(r)}_{ab}}f^{I^{(r)}_{bc}}f^{I^{(r)}_{ca}}. 
\label{eq:intyci}
\end{eqnarray}

For $a$, $b$ and $c$ satisfying 
$M^{(r)}_{ab}M^{(r)}_{bc}M^{(r)}_{ca}>0$ 
(that is equivalent to 
$M^{(r)}_{ac}M^{(r)}_{cb}M^{(r)}_{ba}<0$), 
the overlap integral (\ref{eq:intyci}) is evaluated as 
\begin{eqnarray}
\tilde\lambda^{(r)}_{
I^{(r)}_{ab}I^{(r)}_{bc}I^{(r)}_{ca}} &=& 
\left\{ \begin{array}{lll}
\tilde\lambda^{(r)}_{ab,c} 
&& (M^{(r)}_{ab}>0) \\*[5pt]
\tilde\lambda^{(r)}_{bc,a} 
&& (M^{(r)}_{bc}>0) \\*[5pt]
\tilde\lambda^{(r)}_{ca,b} 
&& (M^{(r)}_{ca}>0) 
\end{array} \right., 
\label{eq:yci}
\end{eqnarray}
where 
\begin{eqnarray}
\tilde\lambda^{(r)}_{ab,c} &=& 
{\cal N}_{M^{(r)}_{ab}}^{-1}
{\cal N}_{M^{(r)}_{bc}}
{\cal N}_{M^{(r)}_{ca}}
\sum_{m=1}^{M^{(r)}_{ab}} 
\delta_{I^{(r)}_{bc}+I^{(r)}_{ca}
-m M^{(r)}_{bc},\, I^{(r)}_{ab}}\, 
\nonumber \\ && \times 
\exp \left[ \frac{\pi i}{{\rm Im}\, \tau_r} 
\left( 
\frac{\bar\zeta^{(r)}_{ab}}{M^{(r)}_{ab}}
{\rm Im}\,\zeta^{(r)}_{ab}
+\frac{\bar\zeta^{(r)}_{bc}}{M^{(r)}_{bc}}
{\rm Im}\,\zeta^{(r)}_{bc}
+\frac{\bar\zeta^{(r)}_{ca}}{M^{(r)}_{ca}}
{\rm Im}\,\zeta^{(r)}_{ca}
\right) \right] 
\nonumber \\ && \times 
\vartheta 
\begin{bmatrix}
\frac{
M^{(r)}_{bc}I^{(r)}_{ca}
-M^{(r)}_{ca}I^{(r)}_{bc}
+m M^{(r)}_{bc}M^{(r)}_{ca}}{
M^{(r)}_{ab}M^{(r)}_{bc}M^{(r)}_{ca}} 
\\ 0 \end{bmatrix} \left( 
\bar\zeta^{(r)}_{ca}M^{(r)}_{bc}
-\bar\zeta^{(r)}_{bc}M^{(r)}_{ca},\, 
-\bar\tau_r M^{(r)}_{ab}M^{(r)}_{bc}M^{(r)}_{ca} 
\right). 
\nonumber
\end{eqnarray}
For $a$, $b$ and $c$ satisfying 
$M^{(r)}_{ab}M^{(r)}_{bc}M^{(r)}_{ca}=0$, 
on the other hand, 
the integral (\ref{eq:intyci}) is given by 
$\tilde\lambda^{(r)}_{
I^{(r)}_{ab}I^{(r)}_{bc}I^{(r)}_{ca}}
= ({\cal A}^{(r)})^{-1/2}$ 
instead of Eq.~(\ref{eq:yci}).

\section{Local supersymmetry and moduli multiplets}
\label{sec:sugra}

The above derivation of 4D effective action has 
been performed in a limit of global supersymmetry, 
because our staring point is 10D SYM theory. 
{}From both theoretical and phenomenological 
viewpoints, however, theories with a local 
supersymmetry are desirable. It is well known that 
10D SYM theories can be embedded into supergravity. 
Actually, low energy effective theories of 
heterotic and type I superstrings as well as 
type II orientifold/D-brane models are categorized 
into such a supergravity-YM system. 

The existence of local supersymmetry implies that 
SYM system is coupled to gravity. With nonvanishing 
gravitational interactions, theories in 
higher-dimensional spacetime in general yield more 
massless modes from gravitational fields which form 
supermultiplets in the 4D effective theories. 
Especially, massless modes originating from the 
extra-dimensional components of higher-dimensional 
tensor (such as 10D graviton fields) and vector fields 
are called moduli which form chiral multiplets in 4D 
${\cal N}=1$ supersymmetry. 

In the following, we assume that there exists a local 
supersymmetry at the starting point of our previous 
analysis. Then, we show how to recover the local 
supersymmetry in the 4D effective theory, and identify 
the dependence of the effective action on geometric 
(so-called closed string) moduli 
as well as dilaton superfields.

\subsection{4D ${\cal N}=1$ effective supergravity}
\label{ssec:4dsugra}

The action for 
4D ${\cal N}=1$ conformal supergravity~\cite{Kaku:1977rk} 
is generally written as\footnote{
The superspace integrals 
$\int d^4 \theta \cdots$ and  $\int d^2 \theta \cdots$ 
in Eq.~(\ref{eq:csugra}) are a kind of simplified expressions 
for a notational convenience, which should be interpreted as 
$D$-term formula $[\ldots]_D$ and 
$F$-term formula $[\ldots]_F$ of 
superconformal tensor calculus~\cite{Kugo:1982cu}, respectively.} 
\begin{eqnarray}
S_{{\cal N}=1} &=& \int d^4x \sqrt{-g^C} \Bigg[ 
-3 \int d^4 \theta\,\bar{C}C \, e^{-K/3} 
\nonumber \\ && \qquad \qquad \qquad 
+\left\{ \int d^2 \theta \left( 
\frac{1}{4} \sum_a f_a\, W^{a,\alpha} W^a_\alpha 
+C^3 W\right) +{\rm h.c.} \right\} \Bigg], 
\label{eq:csugra}
\end{eqnarray}
where $C$ is a compensator chiral superfield, whose 
lowest component $C|_0 \equiv C|_{\theta=\bar\theta=0}$ 
in the $\theta$ and $\bar\theta$ expansion relates the 
4D metric $g^C_{\mu \nu}$ in Eq.~(\ref{eq:csugra}) 
with the one in Einstein frame $g^E_{\mu \nu}$ as 
$g^C_{\mu \nu} = 
(C \bar{C}|_0)^{-1} e^{K|_0/3} g^E_{\mu \nu}$. 
Here and hereafter we denote the lowest component 
of a (function of) superfield $\Phi$ in the $\theta$ and 
$\bar\theta$ expansion as 
$\Phi|_{\theta=\bar\theta=0} \equiv \Phi|_0$. 
The action of Poincare supergravity in Einstein frame 
is obtained~\cite{Kaku:1978ea} 
by a dilatation gauge fixing $C|_0 = e^{K|_0/6}$ 
in the conformal supergravity action~(\ref{eq:csugra}). 
Here and hereafter, we work in a unit that 
the 4D Planck scale is unity. 

When we consider a situation that our starting 10D SYM theory 
is embedded in 10D supergravity, it is important to notify 
that there exists a scalar field $\phi_{10}$ called dilaton 
in 10D supergravity-YM system, and the YM gauge coupling $g$ 
is determined by a vacuum expectation value of 10D dilaton 
field $\phi_{10}$ as 
\begin{eqnarray}
g &=& e^{\vev{\phi_{10}}/2}. 
\nonumber
\end{eqnarray}
Furthermore, the 4D effective action (\ref{eq:seff}) 
is written in a so-called string frame, which is obtained 
by Eq.~(\ref{eq:csugra}) with a dilatation gauge fixing, 
\begin{eqnarray}
C|_0 &=& e^{-\phi_4} e^{K|_0/6}, 
\label{eq:stf}
\end{eqnarray}
where $\phi_4$ is a zero-mode of $\phi_{10}$, 
which we call a 4D dilaton field. 
The VEV of $\phi_4$ satisfies 
\begin{eqnarray}
e^{-2 \vev{\phi_4}} &=& 
e^{-2 \vev{\phi_{10}}} \prod_i {\cal A}^{(i)} 
\ = \ g^{-4} \prod_i {\cal A}^{(i)}. 
\label{eq:phi4vev}
\end{eqnarray}
The YM-field independent part of the K\"ahler potential $K$ 
in the action (\ref{eq:csugra}), that we denote $K^{(0)}$, 
is well known as 
\begin{eqnarray}
K^{(0)} &=& 
\ln \frac{g^8}{2^7 \left( 
\prod_i {\cal A}_i \right)^2 
\prod_i {\rm Im}\, \tau_i}, 
\nonumber
\end{eqnarray}
determined from a coefficient of 
Einstein-Hilbert term in dimensionally reduced 
4D effective supergravity action. 

In the string frame (\ref{eq:stf}), we expand the action 
(\ref{eq:csugra}) in powers of YM-fields and compare the 
corresponding terms in the power series to those in globally 
supersymmetric effective action (\ref{eq:seff}). 
In this way, we determine the K\"ahler potential $K$, 
the superpotential $W$ and the gauge kinetic function $f_a$ 
in the 4D effective supergravity action~(\ref{eq:csugra}) as 
\begin{eqnarray}
K &=& K^{(0)} + \sum_{i,j}\,\sum_{a,b}\,
\sum_{{\cal I}_{ab}} Z_{{\cal I}_{ab}}^{\bar{i}j} 
{\rm tr} \left[ \bar\phi_{\bar{i}}^{{\cal I}_{ab}} e^{-V^a} 
\phi_j^{{\cal I}_{ab}} e^{V^b} \right], 
\label{eq:ksugra} \\
W &=& 
\sum_{i,j,k}\,\sum_{a,b,c}\,
\sum_{{\cal I}_{ab},\,
{\cal I}_{bc},\,
{\cal I}_{ca}}
\lambda_{{\cal I}_{ab}
{\cal I}_{bc}
{\cal I}_{ca}}^{ijk} 
{\rm tr} \left[ 
\phi_i^{{\cal I}_{ab}} 
\phi_j^{{\cal I}_{bc}} 
\phi_k^{{\cal I}_{ca}} \right], 
\label{eq:wsugra} \\
f_a &=& \frac{1}{g^2} \prod_i {\cal A}^{(i)}, 
\label{eq:fsugra}
\end{eqnarray}
where 
the K\"ahler metric 
$Z_{{\cal I}_{ab}}^{\bar{i}j}$ 
and the holomorphic Yukawa couplings 
$\lambda_{{\cal I}_{ab}
{\cal I}_{bc}
{\cal I}_{ca}}^{ijk}$ 
are given by 
\begin{eqnarray}
Z_{{\cal I}_{ab}}^{\bar{i}j} &=& 
e^{2 \vev{\phi_4}} 
\tilde{Z}_{{\cal I}_{ab}}^{\bar{i}j}, 
\label{eq:kmsugra} \\
\lambda_{{\cal I}_{ab}
{\cal I}_{bc}
{\cal I}_{ca}}^{ijk} 
&=& e^{3 \vev{\phi_4}} e^{-K^{(0)}/2}
\tilde\lambda_{{\cal I}_{ab}
{\cal I}_{bc}
{\cal I}_{ca}}^{ijk}, 
\label{eq:ycsugra}
\end{eqnarray}
and 
$\tilde{Z}_{{\cal I}_{ab}}^{\bar{i}j}$, 
$\tilde\lambda_{{\cal I}_{ab}
{\cal I}_{bc}{\cal I}_{ca}}^{ijk}$
are given in Eqs.~(\ref{eq:km}) and (\ref{eq:yc}), 
respectively. 
Note that the VEV of 4D dilaton field $\vev{\phi_4}$ 
in Eqs.~(\ref{eq:kmsugra}) and (\ref{eq:ycsugra}) 
is related to $g$ and ${\cal A}^{(i)}$ 
as shown in Eq.~(\ref{eq:phi4vev}).

\subsection{Moduli dependence}
\label{ssec:moduli}
As mentioned above, the YM gauge coupling constant $g$ 
is determined by a VEV of 10D dilaton field $\phi_{10}$ 
which is a dynamical scalar field. 
Furthermore, the geometric (torus) parameters 
$R_i$ and $\tau_i$ in the metric (\ref{eq:gtorus}) should 
be considered as VEVs of dynamical fields originating from 
10D metric $G_{MN}$. The zero-modes of these dynamical 
fields are called moduli. 
In the case of toroidal compactification, there exist 
K\"ahler moduli and complex structure moduli whose VEVs 
determine $R_i$ and $\tau_i$, respectively, in addition 
to the zero-mode $\phi_4$ of 10D dilaton field $\phi_{10}$. 

We assume (at least ${\cal N}=1$) supersymmetric vacuum 
configurations (\ref{eq:vevs}) in this paper, and then 
geometric moduli as well as $\phi_4$ form ${\cal N}=1$ 
supermultiplets described by chiral superfields. 
Because these moduli fields are singlets under gauge 
transformations of SYM sector, they do not feel magnetic 
fluxes. Therefore, the formation of moduli supermultiplets 
is the same one as a pure toroidal case without magnetic 
fluxes, where we usually denote the dilaton, K\"ahler and 
complex-structure moduli chiral superfields as 
$S$, $T_i$ and $U_i$, respectively. 
A suitable identification of 
the VEVs of these fields are known~\cite{Cremades:2004wa} as
\begin{eqnarray}
{\rm Re}\,\vev{S}|_0 &\equiv& g^{-2} \prod_i {\cal A}^{(i)}, 
\qquad 
{\rm Re}\,\vev{T_i}|_0 \ \equiv \ g^{-2} {\cal A}^{(i)}, 
\qquad 
\vev{U_i}|_0 \ \equiv \ i \bar\tau_i. 
\label{eq:stuvev}
\end{eqnarray}

An important task here is to identify 
the dependence of 4D ${\cal N}=1$ effective supergravity 
action (\ref{eq:csugra}) on the dilaton and moduli chiral 
superfields $S$, $T_i$ and $U_i$. The K\"ahler and 
superpotential as well as the gauge kinetic functions 
in the action (\ref{eq:csugra}) are shown in 
Eqs.~(\ref{eq:ksugra}), (\ref{eq:wsugra}) and 
(\ref{eq:fsugra}), which are written in terms of VEVs 
of moduli, namely, parameters $g$, $R_i$ and $\tau_i$. 
Correct combinations of these parameters should be 
promoted to moduli chiral superfields $S$, $T_i$ and $U_i$ 
following the relations (\ref{eq:stuvev}) 
up to a certain parameter-dependent rescaling 
of YM superfields $V^a$ and $\phi_i^{{\cal I}_{ab}}$ 
in the action (\ref{eq:csugra}). 
The rescaling is performed as 
\begin{eqnarray}
V^a &\to& V^a, \qquad 
\phi_i^{{\cal I}_{ab}} \ \to \ 
\alpha^{(i)}_{ab} \phi_i^{{\cal I}_{ab}}, 
\label{eq:rescale}
\end{eqnarray}
where, 
for $\prod_r M^{(r)}_{ab} M^{(r)}_{bc} M^{(r)}_{ca} \ne 0$, 
\begin{eqnarray}
\alpha^{(i)}_{ab} &=& 
\frac{1}{g \sqrt{2\,{\rm Im}\,\tau_i}}\, 
\left( \prod_r 
\frac{{\cal A}^{(r)}}{\sqrt{2\,{\rm Im}\,\tau_r}} \right)^{1/2} 
\exp \left[ -\sum_r \frac{\pi i}{{\rm Im}\,\tau_r} 
\frac{\bar\zeta^{(r)}_{ab}}{M^{(r)}_{ab}} 
{\rm Im}\,\zeta^{(r)}_{ab} \right]
\left( \frac{|M^{(i)}_{ab}|}{
\prod_{r \ne i} |M^{(r)}_{ab}|} \right)^{1/4}, 
\label{eq:fred}
\end{eqnarray}
so that all the parameters $g$, $R_i$ and $\tau_i$ 
can be promoted to superfields $S$, $T_i$ and $U_i$ 
through Eq.~(\ref{eq:stuvev}) 
with a proper holomorphicity in the superspace action. 
In a case with some vanishing fluxes, 
$\prod_r M^{(r)}_{ab} M^{(r)}_{bc} M^{(r)}_{ca}=0$, 
the way of rescaling is shown in Appendix~\ref{sec:app}. 

After the above rescaling and promotion 
for $\prod_r M^{(r)}_{ab} M^{(r)}_{bc} M^{(r)}_{ca} \ne 0$, 
we find moduli dependence of 
the YM-field independent part of the 
K\"ahler potential $K^{(0)}$, 
the K\"ahler metric 
$Z_{{\cal I}_{ab}}^{\bar{i}j}$, 
the holomorphic Yukawa couplings 
$\lambda_{{\cal I}_{ab}
{\cal I}_{bc}{\cal I}_{ca}}^{ijk}$ 
and the gauge kinetic functions $f_a$ as 
\begin{eqnarray}
K^{(0)}(S,T,U) &=& 
-\ln (S+\bar{S}) 
-\sum_r \ln(T_r + \bar{T}_r)  
-\sum_r \ln(U_r + \bar{U}_r), 
\label{eq:k0mod} \\
Z_{{\cal I}_{ab}}^{\bar{i}j}(S,T,U) 
&=& \delta^{\bar{i}j} 
\left( \frac{T_j+\bar{T}_{\bar{j}}}{2} \right)^{-1} 
\left( \prod_r 
\frac{U_r+\bar{U}_{\bar{r}}}{2} \right)^{-1/2} 
\nonumber \\ && \times 
\frac{1}{2^{5/2}}
\left( \frac{|M^{(j)}_{ab}|} 
{\prod_{r \ne j} |M^{(r)}_{ab}|} \right)^{1/2} 
\exp \left[ -\sum_r \frac{4 \pi}{U_r+\bar{U}_{\bar{r}}} 
\frac{({\rm Im}\,\zeta^{(r)}_{ab})^2}{M^{(r)}_{ab}} \right], 
\label{eq:kmmod} \\
\lambda_{{\cal I}_{ab}
{\cal I}_{bc}{\cal I}_{ca}}^{ijk}(U) 
&=& 
-\frac13\epsilon^{{\rm i}{\rm j}{\rm k}} 
\delta_{{\rm i}}^{\ i} 
\delta_{{\rm j}}^{\ j} 
\delta_{{\rm k}}^{\ k} 
\prod_r 
\lambda^{(r)}_{
I^{(r)}_{ab}I^{(r)}_{bc}I^{(r)}_{ca}}(U),  
\label{eq:ycmod} \\
f_a(S) &=& S. 
\label{eq:famod}
\end{eqnarray}
For $a$, $b$ and $c$ satisfying 
$M^{(r)}_{ab}M^{(r)}_{bc}M^{(r)}_{ca}>0$ 
(that is equivalent to 
$M^{(r)}_{ac}M^{(r)}_{cb}M^{(r)}_{ba}<0$), 
the function 
$\lambda^{(r)}_{
I^{(r)}_{ab}I^{(r)}_{bc}I^{(r)}_{ca}}(U)$ 
in Eq.~(\ref{eq:ycmod}) is evaluated as 
\begin{eqnarray}
\lambda^{(r)}_{
I^{(r)}_{ab}I^{(r)}_{bc}I^{(r)}_{ca}}(U) &=& 
\left\{ \begin{array}{lll}
\lambda^{(r)}_{ab,c}(U) 
&& (M^{(r)}_{ab}>0) \\*[5pt]
\lambda^{(r)}_{bc,a}(U) 
&& (M^{(r)}_{bc}>0) \\*[5pt]
\lambda^{(r)}_{ca,b}(U) 
&& (M^{(r)}_{ca}>0) 
\end{array} \right., 
\label{eq:ycimod}
\end{eqnarray}
where 
\begin{eqnarray}
\lambda^{(r)}_{ab,c}(U) &=& 
\sum_{m=1}^{M^{(r)}_{ab}} 
\delta_{I^{(r)}_{bc}+I^{(r)}_{ca}
-m M^{(r)}_{bc},\, I^{(r)}_{ab}}\, 
\nonumber \\ && \times 
\vartheta 
\begin{bmatrix}
\frac{
M^{(r)}_{bc}I^{(r)}_{ca}
-M^{(r)}_{ca}I^{(r)}_{bc}
+m M^{(r)}_{bc}M^{(r)}_{ca}}{
M^{(r)}_{ab}M^{(r)}_{bc}M^{(r)}_{ca}} 
\\ 0 \end{bmatrix} \left( 
\bar\zeta^{(r)}_{ca}M^{(r)}_{bc}
-\bar\zeta^{(r)}_{bc}M^{(r)}_{ca},\, 
iU_r M^{(r)}_{ab}M^{(r)}_{bc}M^{(r)}_{ca} 
\right). 
\nonumber
\end{eqnarray}
In a case with some vanishing fluxes, 
$\prod_r M^{(r)}_{ab} M^{(r)}_{bc} M^{(r)}_{ca}=0$, 
the expression of K\"ahler metric is shown 
in Appendix~\ref{sec:app}. 

Note that these functions of moduli 
(\ref{eq:k0mod})-(\ref{eq:famod}) 
appear in the action (\ref{eq:csugra}) with 
the K\"ahler potential (\ref{eq:ksugra}) and 
the superpotential (\ref{eq:wsugra}) where 
the YM fields $V^a$ and $\phi_i^{{\cal I}_{ab}}$ 
represent those after the rescaling (\ref{eq:fred}). 
These results (\ref{eq:k0mod})-(\ref{eq:famod}) 
are consistent with those obtained in different 
ways~\cite{Cremades:2004wa,DiVecchia:2008tm}. 
A systematic formulation of 4D effective theory 
respecting ${\cal N}=1$ superspace structure 
presented here will be easily adopted to more 
general systems of magnetized SYM theories 
and D-branes.

\section{An example of model building}
\label{sec:model}
In this section, we indicate a possible direction of 
phenomenological model building based on our formulation. 
Starting from 10D $U(N)$ SYM theory with $N=8$, we assume 
magnetic fluxes yielding $\tilde{N}=3$ and 
$(N_1,N_2,N_3) \equiv (N_C,N_L,N_R)=(4,2,2)$ 
in Eq.~(\ref{eq:gsb}) that break YM symmetry 
as $U(8) \to U(4)_C \times U(2)_L \times U(2)_R$. 
We consider the case that this is further broken down to 
$U(3)_C \times U(1)_{C'} \times U(2)_L 
\times U(1)_{R'} \times U(1)_{R''}$
by Wilson-lines yielding 
$\tilde{N}=5$ and $(N_1,N_2,N_3,N_4,N_5) 
\equiv (N_C,N_{C'},N_L,N_{R'},N_{R''}) = (3,1,2,1,1)$. 
The situation is realized~\cite{Ohki:2010nf} 
by the following magnetic fluxes and Wilson-lines 
for $r=1,2,3$: 
\begin{eqnarray}
F_{2+2r,3+2r} &=& 2 \pi 
\begin{pmatrix}
M^{(r)}_C {\bm 1}_4 & \ & \ \\
\ & M^{(r)}_L {\bm 1}_2 & \ \\
\ & \ & M^{(r)}_R  {\bm 1}_2 
\end{pmatrix}, 
\nonumber \\
\zeta_r &=& 
\begin{pmatrix}
\zeta^{(r)}_C {\bm 1}_3 & \ & \ & \ & \ \\
\ & \zeta^{(r)}_{C'} & \ & \ & \ \\
\ & \ & \zeta^{(r)}_L {\bm 1}_2 & \ & \ \\
\ & \ & \ & \zeta^{(r)}_{R'} & \ \\
\ & \ & \ & \ & \zeta^{(r)}_{R''} 
\end{pmatrix}, 
\nonumber
\end{eqnarray}
where ${\bm 1}_N$ is a $N \times N$ unit matrix, and 
all the nonvanishing entries take different values 
from each other. 

We embed the gauge symmetries $SU(3)_C$ and $SU(2)_L$ 
of the standard model into the above unbroken gauge groups 
as $SU(3)_C \subset U(3)_C$ and $SU(2)_L \subset U(2)_L$. 
Then, in order to obtain three generations of quarks and 
leptons from the zero-mode degeneracy (\ref{eq:igen}) and 
full-rank Yukawa matrices from the 10D gauge interaction, 
the magnetic fluxes are determined, e.g., as 
\begin{eqnarray}
(M^{(1)}_C,M^{(1)}_L,M^{(1)}_R) &=& (0,+3,-3), 
\nonumber \\
(M^{(2)}_C,M^{(2)}_L,M^{(2)}_R) &=& (0,-1,0), 
\nonumber \\
(M^{(3)}_C,M^{(3)}_L,M^{(3)}_R) &=& (0,0,+1), 
\label{eq:model}
\end{eqnarray}
which correspond to 
\begin{eqnarray}
M^{(1)}_C-M^{(1)}_L \ = \ -3, \qquad 
M^{(1)}_L-M^{(1)}_R &=& +6, \qquad 
M^{(1)}_R-M^{(1)}_C \ = \ -3, \nonumber \\
M^{(2)}_C-M^{(2)}_L \ = \ +1, \qquad 
M^{(2)}_L-M^{(2)}_R &=& -1, \qquad 
M^{(2)}_R-M^{(2)}_C \ = \ 0, \nonumber \\
M^{(3)}_C-M^{(3)}_L \ = \ 0, \qquad 
M^{(3)}_L-M^{(3)}_R &=& -1, \qquad 
M^{(3)}_R-M^{(3)}_C \ = \ +1. 
\label{eq:f3gen}
\end{eqnarray}
In this case, supersymmetry conditions 
(\ref{eq:abd}) and (\ref{eq:abfi}) are satisfied by 
\begin{eqnarray}
{\cal A}^{(1)}/{\cal A}^{(2)}
={\cal A}^{(1)}/{\cal A}^{(3)}=3. 
\label{eq:susy}
\end{eqnarray}

In this model, chiral superfields 
$Q$, $U$, $D$, $L$, $N$, $E$, $H_u$ and $H_d$ 
carrying 
the left-handed quark ($Q$), 
the right-handed up-type quark ($U$), 
the right-handed down-type quark ($D$), 
the left-handed lepton ($L$), 
the right-handed neutrino ($N$), 
the right-handed electron ($E$), 
the up-type Higgs particle ($H_u$) and 
the down-type Higgs particle ($H_d$), respectively, 
are found in $\phi_i^{{\cal I}_{ab}}$ as 
\begin{eqnarray}
\phi_1^{{\cal I}_{ab}} &=& 
\left( 
\begin{array}{cc|c|cc}
\Omega_C^{(1)} & \Xi_{CC'}^{(1)} & 0 & 
\Xi_{CR'}^{(1)} & \Xi_{CR''}^{(1)} \\
\Xi_{C'C}^{(1)} & \Omega_{C'}^{(1)} & 0 & 
\Xi_{C'R'}^{(1)} & \Xi_{C'R''}^{(1)} \\ 
\hline 
\Xi_{LC}^{(1)} & \Xi_{LC'}^{(1)} & \Omega_L^{(1)} & 
H_u^K & H_d^K \\ 
\hline 
0 & 0 & 0 & \Omega_{R'}^{(1)} & \Xi_{R'R''}^{(1)} \\
0 & 0 & 0 & \Xi_{R''R'}^{(1)} & \Omega_{R''}^{(1)} 
\end{array}
\right), 
\label{eq:phi1cont} \\
\phi_2^{{\cal I}_{ab}} &=& 
\left( 
\begin{array}{cc|c|cc}
\Omega_C^{(2)} & \Xi_{CC'}^{(2)} & Q^I & 0 & 0 \\
\Xi_{C'C}^{(2)} & \Omega_{C'}^{(2)} & L^I & 0 & 0 \\
\hline 
0 & 0 & \Omega_L^{(2)} & 0 & 0 \\
\hline 
0 & 0 & 0 & \Omega_{R'}^{(2)} & \Xi_{R'R''}^{(2)} \\
0 & 0 & 0 & \Xi_{R''R'}^{(2)} & \Omega_{R''}^{(2)} 
\end{array}
\right), 
\label{eq:phi2cont} \\
\phi_3^{{\cal I}_{ab}} &=& 
\left( 
\begin{array}{cc|c|cc}
\Omega_C^{(3)} & \Xi_{CC'}^{(3)} & 0 & 0 & 0 \\
\Xi_{C'C}^{(3)} & \Omega_{C'}^{(3)} & 0 & 0 & 0 \\
\hline 
0 & 0 & \Omega_L^{(3)} & 0 & 0 \\
\hline 
U^J & N^J & 0 & \Omega_{R'}^{(3)} & \Xi_{R'R''}^{(3)} \\
D^J & E^J & 0 & \Xi_{R''R'}^{(3)} & \Omega_{R''}^{(3)} 
\end{array}
\right), 
\label{eq:phi3cont}
\end{eqnarray}
where the rows and columns of matrices 
correspond to $a=1,\ldots,5$ and $b=1,\ldots,5$, 
respectively, and the indices $I,J=1,2,3$ and 
$K=1,\ldots,6$ label generations. 
Three generations of 
$Q$, $U$, $D$, $L$, $N$, $E$ and 
six generations of $H_u$ and $H_d$ 
are generated by the fluxes~(\ref{eq:model}). 
The K\"ahler metric and holomorphic Yukawa couplings 
for these superfields are easily derived from 
Eqs.~(\ref{eq:kmmod}) and (\ref{eq:ycmod}). 

Note that each of zero entries in the matrices 
(\ref{eq:phi1cont}), (\ref{eq:phi2cont}) and 
(\ref{eq:phi3cont}) represents eliminated components 
due to the effect of chirality projection caused 
by magnetic fluxes. However, because we require 
some vanishing fluxes in Eq.~(\ref{eq:f3gen}) 
in order to obtain three generations of quarks and 
leptons, there appear some massless exotic modes 
$\Xi_{ab}^{(r)}$ as well as diagonal components 
$\Omega_{a}^{(r)}$ (so-called open string moduli), 
all of which feel zero fluxes. 
Some of these modes can be eliminated 
if we consider certain orbifold projections on 
$r=2,3$ tori, that is, 
a magnetized orbifold model~\cite{Abe:2008fi}. 
More details of this model building and 
phenomenological features at a low energy 
will be reported in a separate paper~\cite{akos}.

\section{Conclusion}
\label{sec:conc}
We have presented 4D ${\cal N}=1$ superfield description 
of 10D SYM theories compactified on magnetized tori which 
preserve the ${\cal N}=1$ supersymmetry. Based on such a 
description, we have derived 4D effective action for 
massless zero-modes written in the ${\cal N}=1$ superspace. 
We further identified moduli dependence of the effective 
action by promoting the YM gauge coupling constant $g$ and 
geometric parameters $R_i$ and $\tau_i$ to a dilaton, 
K\"ahler and complex-structure moduli superfields. 
The resulting effective supergravity action would be 
useful for building phenomenological models 
and for analyzing them systematically. 

Although we have worked on 10D SYM theories in this paper, 
it is straightforward to adopt our formulation to SYM in 
lower-than-ten dimensional spacetime, in a similar way to 
the one suggested in Ref.~\cite{ArkaniHamed:2001tb} without 
magnetic fluxes. A local supersymmetry can be recovered 
in 4D effective theories\footnote{
The corresponding supergravity actions in the original 
higher-dimensional spacetime could be also written in 
${\cal N}=1$ superspace as shown in the case of 
five dimensions~\cite{Paccetti:2004ri}.} 
following the procedure presented in Sec.~\ref{sec:sugra}. 
Then, e.g., in type IIB orientifolds, 
our formulation will be applied not only to magnetized D9 
branes (a class of which is T-dual to intersecting D6 branes 
in IIA side), but also to D5-D9~\cite{DiVecchia:2011mf} and 
D3-D7 brane configurations with magnetic fluxes in extra dimensions. 
Lower-dimensional brane configurations may allow the introduction of 
supersymmetry-breaking branes sequestered from the visible sector, 
which coincide with flavor structures in the visible sector 
generated by magnetic fluxes. 

The explicit moduli dependence of the superspace effective 
action also allows us to study a moduli stabilization and a 
supersymmetry breaking at a Minkowski minimum~\cite{Kachru:2003aw} 
based on SYM theories, by minimizing the moduli and hidden-sector 
potential generated by some combinations~\cite{Dudas:2006gr} of 
nonperturbative effects and 
a dynamical supersymmetry breaking~\cite{Intriligator:2006dd}. 
Then, it would be possible to determine explicit forms of 
soft terms in the visible sector generated by moduli-mediated 
supersymmetry breaking (or mixed modulus-anomaly 
mediation~\cite{Choi:2004sx}). 
In such models, brane configurations in the higher-dimensional 
spacetime might be detected by measuring supersymmetric flavor 
structures at a low energy. 
The formulation of 4D effective action presented here 
would be suitable for such analyses.

\subsection*{Acknowledgement}
The work of H.~A. was supported by the Waseda University 
Grant for Special Research Projects No.2011B-177. 
The work of T.~K. is supported in part by a Grant-in-Aid for 
Scientific Research No.~20540266 and the Grant-in-Aid for 
the Global COE Program ``The Next Generation of Physics, 
Spun from Universality and Emergence'' from the Ministry of 
Education, Culture, Sports, Science and Technology of Japan. 
The work of H.~O. is supported by the JSPS Grant-in-Aid for 
Scientific Research (S) No. 22224003. 
H.~A. and T.~K. thank the Yukawa Institute for Theoretical 
Physics at Kyoto University. 
Discussions during the YITP workshop~"Summer Institute 2011" 
were useful to complete this work.

\appendix

\section{Rescaling fields with some vanishing fluxes}
\label{sec:app}

In typical cases with some vanishing fluxes, 
$\prod_r M^{(r)}_{ab} M^{(r)}_{bc} M^{(r)}_{ca}=0$, 
the way of rescaling (\ref{eq:rescale}) and 
the corresponding K\"ahler metric are found as 
\begin{eqnarray}
\alpha^{(i)}_{ab} &=& 
\frac{1}{g \sqrt{2\,{\rm Im}\,\tau_i}}\, 
\left( \prod_r 
\frac{{\cal A}^{(r)}}{\sqrt{2\,{\rm Im}\,\tau_r}} \right)^{1/2} 
\exp \left[ -\sum_{r \ne k} \frac{\pi i}{{\rm Im}\,\tau_r} 
\frac{\bar\zeta^{(r)}_{ab}}{M^{(r)}_{ab}} 
{\rm Im}\,\zeta^{(r)}_{ab} \right]
\nonumber \\ && \times 
\left( \frac{|M^{(i)}_{ab}|}{
\prod_{r \ne i} |M^{(r)}_{ab}|} \right)^{1/4} 
\left( 2 {\rm Im}\,\tau_k |M^{(k)}_{ab}| \right)^{1/4}, 
\nonumber
\end{eqnarray}
and 
\begin{eqnarray}
Z_{{\cal I}_{ab}}^{\bar{i}j}(S,T,U) 
&=& \delta^{\bar{i}j} 
\left( \frac{T_j+\bar{T}_{\bar{j}}}{2} \right)^{-1} 
\left( \prod_{r \ne k} 
\frac{U_r+\bar{U}_{\bar{r}}}{2} \right)^{-1/2} 
\nonumber \\ && \times 
\frac{1}{2^{2}} \left( \frac{|M^{(j)}_{ab}|} 
{\prod_{r \ne j} |M^{(r)}_{ab}|} \right)^{1/2} |M^{(k)}_{ab}|^{1/2}
\exp \left[ -\sum_{r \ne k} \frac{4 \pi}{U_r+\bar{U}_{\bar{r}}} 
\frac{({\rm Im}\,\zeta^{(r)}_{ab})^2}{M^{(r)}_{ab}} \right], 
\nonumber
\end{eqnarray}
for $M^{(k)}_{ab}=0$ and $^\exists k \ne i$ 
with others nonvanishing, and 
\begin{eqnarray}
\alpha^{(i)}_{ab} &=& 
\frac{1}{g \sqrt{2\,{\rm Im}\,\tau_i}}\, 
\left( \prod_r 
\frac{{\cal A}^{(r)}}{\sqrt{2\,{\rm Im}\,\tau_r}} \right)^{1/2} 
\exp \left[ -\sum_{r \ne k} \frac{\pi i}{{\rm Im}\,\tau_r} 
\frac{\bar\zeta^{(r)}_{ab}}{M^{(r)}_{ab}} 
{\rm Im}\,\zeta^{(r)}_{ab} \right]
\nonumber \\ && \times 
\left( \frac{|M^{(i)}_{ab}|}{
\prod_{r \ne i} |M^{(r)}_{ab}|} \right)^{1/4} 
|M^{(k)}_{ab}|^{(-1)^{\delta_{ik}}/4}, 
\nonumber
\end{eqnarray}
and 
\begin{eqnarray}
Z_{{\cal I}_{ab}}^{\bar{i}j}(S,T,U) 
&=& \delta^{\bar{i}j} 
\left( \frac{T_j+\bar{T}_{\bar{j}}}{2} \right)^{-1} 
\left( \prod_r 
\frac{U_r+\bar{U}_{\bar{r}}}{2} \right)^{-1/2} 
\nonumber \\ && \times 
\frac{1}{2^{5/2}} \left( \frac{|M^{(j)}_{ab}|} 
{\prod_{r \ne j} |M^{(r)}_{ab}|} \right)^{1/2} 
|M^{(k)}_{ab}|^{(-1)^{\delta_{jk}}/2}
\exp \left[ -\sum_{r \ne k} \frac{4 \pi}{U_r+\bar{U}_{\bar{r}}} 
\frac{({\rm Im}\,\zeta^{(r)}_{ab})^2}{M^{(r)}_{ab}} \right], 
\nonumber
\end{eqnarray}
for $M^{(k)}_{bc}=0$ or $M^{(k)}_{ca}=0$ and $^\exists k$ 
with others nonvanishing, and so on.

\end{document}